\documentclass[useAMS]{mn2e}
\usepackage{times}
\input{epsf}

\title[Scaling black hole variability]
{Scaling variability from stellar to supermassive black holes}

\author[C.~Done and M.~Gierli{\'n}ski ] {Chris Done$^{1}$ and
Marek~Gierli\'nski$^{1,2}$ \\
$^1$Department of Physics, University of Durham, South Road, Durham
DH1 3LE, UK\\
$^2$Obserwatorium Astronomiczne Uniwersytetu Jagiello{\'n}skiego, 30-244
Krak{\'o}w, Orla 171, Poland}

\date{Submitted to MNRAS}
\pagerange{\pageref{firstpage}--\pageref{lastpage}}
\pubyear{2005}

\begin{document}

\def\aap{A\&A}
\def\apj{ApJ}
\def\apjl{ApJ}
\def\mnras{MNRAS}

\topmargin = -0.5cm

\label{firstpage}

\maketitle

\begin{abstract}

We investigate the correspondence between the variability seen in the
stellar and supermassive black holes. Galactic Black Hole (GBH) power
density spectra (PDS) are generally complex, and dependent on spectral
state. In the low/hard state the high-frequency rollover in the PDS
{\em moves} in a way which is not simply related to luminosity. Hence
this feature can only be used as an approximate indicator rather than
as an accurate tracer of black hole mass in AGN. The X-ray spectrum in
the high/soft state is dominated by the disc in GBH, which is rather
stable. We show that the PDS of the Comptonized tail in GBH can be much
more variable, and that it is this which should be compared to AGN due
to their much lower disc temperature. This bandpass effect removes a
problem in interpreting the (often highly variable) Narrow Line Seyfert
1 (NLS1) galaxies as the counterparts of the high mass accretion rate
GBH. Additionally, we speculate that some NLS1 (e.g. Akn 564) are
counterparts of the very high state. The Comptonized tail in this state
is also highly variable, but with PDS which can be roughly described as
band-limited noise. This shape is similar to that seen in the low/hard
state, so merely seeing such band-limited noise in the power spectrum
of an AGN does {\em not} necessarily imply low luminosity. We also
stress that Cyg X-1, often used for comparison with AGN, is not a
typical black hole system due to its persistent nature. In particular,
the shape of its power spectrum in the high/soft state is markedly
different from that of other (transient) GBH systems in this state. The
fact that the NLS1s NGC~4051 and MCG-6-30-15 do appear to show a power
spectrum similar to that of the high/soft state of Cyg X-1 may lend
observational support to theoretical speculation that the Hydrogen
ionization disc instability does not operate in AGN.

\end{abstract}


\section{Introduction}

Black holes have no hair; they are the simplest celestial objects,
fully characterized by their mass and spin. Scaling between
supermassive and Galactic black hole (GBH) systems should {\em
theoretically} be very simple. There should be a fundamental similarity
between the accretion flow as a function of $L/L_{\rm Edd}$, with only
a weak dependence on black hole mass. For example, standard optically
thick accretion disc models predict that the spectrum should be
dominated by a multi-temperature blackbody component (Shakura \&
Sunyaev 1973), irrespective of the factor of $10^5$--$10^8$ difference
in mass. The only significant difference should be the energy at which
the peak in the spectrum is observed (in $E F_E$ representation, where
$F_E$ is energy flux per unit energy). At the Eddington luminosity this
should be $\sim$1~keV in GBH and $\sim$10~eV in AGN. The idea of a
(mostly) mass-invariant accretion flow is also backed by observational
evidence for simple scaling relations of the accretion flow between
very different mass black holes (Merloni, Heinz \& di Matteo 2003;
Falcke, K{\"o}rding \& Markoff 2004).

However, observations also clearly show that the accretion flow is much
more complicated than the simple Shakura-Sunyaev disc model. GBH show a
variety of different types of X-ray spectral and variability behaviour,
which are used to classify the accretion flow into different spectral
states (e.g. Tanaka \& Lewin 1995; van der Klis 2000; McClintock \&
Remillard 2005; Done \& Gierli{\'n}ski 2003; Zdziarski \&
Gierli{\'n}ski 2004). At high luminosity ($L/L_{\rm Edd}\ga 0.2$) the
GBH show two main spectral states, the high/soft (or thermal dominant)
and very high (steep power law). Both of these generally have peak
energy output at $\sim 1$~keV, as expected from an optically thick disc
at these high mass accretion rates.  However, they also show a tail of
emission to higher energies produced by Compton scattering of seed
photons from the disc by energetic electrons. This tail is rather
complex (e.g. Gierli{\'n}ski et al. 1999), but can roughly be described
by a power law. In the high/soft state this tail is rather weak and has
energy spectral index $\alpha\sim 1$ (where $F_E\propto E^{- \alpha}$)
while in the very high state, the tail carries a large fraction of the
bolometric luminosity and has $\alpha\sim 1.5$.  Conversely, at lower
luminosities the GBH can show qualitatively different emission, forming
the low/hard state.  Here they have only a weak disc component and have
a dominant {\em hard} Comptonized tail.  Again its shape is rather
complex (e.g. the review by Zdziarski \& Gierli{\'n}ski 2004), but can
be roughly described as a power law of energy index $\alpha$ = 0.5--1
(Tanaka \& Lewin 1995; McClintock \& Remillard 2005).

By analogy, the AGN should also show these different X-ray spectral
states (see e.g.  White, Fabian \& Mushotzky 1984 for an early example
of this), and hopefully these intrinsic differences in spectra could
explain some of the different types of AGN behaviour which cannot be
modelled in simple orientation dependent unification schemes. Very
broadly, LINERS have very low $L/L_{\rm Edd}$, so might correspond to
the extreme (hardest spectra) low/hard states seen in GBH, many Seyfert
1's are at a few percent of Eddington, so could be bright low/hard
state, while PG Quasars and especially Narrow Line Seyfert 1's (NLS1's)
are generally at high $L/L_{\rm Edd}$, so would correspond to the
high/soft or very high states of GBH (Pounds, Done \& Osborne 1995; see
e.g. Woo \& Urry 2002, Boronson 2002 and Collin \& Kawaguchi 2004 for
estimates of $L/L_{\rm Edd}$). However, unlike the GBH, transitions
between these states are unobservable, making them more difficult to
identify. The GBH make transitions between the spectral states on
timescales of a few days (e.g. Cui et al. 1997; Wilson \& Done 2001;
Kalemci et al. 2004). This translates to timescales of 300 to
$3\times10^{5}$ years when scaled from $10$ M$_\odot$ to $10^6$--$10^9$
M$_\odot$ black hole mass. The one exception is the very luminous
system GRS 1915+105, where state transitions can occur on timescales of
a few seconds (Belloni et al. 2000), implying 10--1000 days timescales
in supermassive black holes.

Nonetheless, AGN do show substantial variability in their X-ray
emission on timescales of days and even hours (e.g. Nandra et al.
1997).  Instead of true state transitions, this must correspond to the
GBH variability seen on much shorter timescales, between milliseconds
and seconds. This range of timescales is usually studied via power
density spectra (PDS). In the GBH the PDS shape between 0.001--100~Hz
is a strong function of the spectral state of the source, showing that
the spectral and timing behaviour are generally correlated. For
example, in Cyg X-1, the overall shape of the PDS in the low/hard state
can be approximately described as band- limited noise in $\nu P_\nu$
representation (where $\nu$ is frequency and $P_\nu$ is variability
power at that frequency), with a `flat top' ($P_\nu \propto \nu^{- 1}$)
so that most of the power is emitted between two break frequencies,
$\nu_{\rm low}\sim 0.3$~Hz and $\nu_{\rm high}\sim 3$~Hz (e.g.
Gilfanov, Churazov \& Revnivtsev 1999). By contrast, its high/soft
state has a high-frequency break at $\sim 14$~Hz (e.g. Gilfanov et al.
1999) and no low frequency break, so its overall shape is
characteristic of a low-pass filter rather than of band-limited noise.

Here we look in more detail at the correspondence between AGN and GBH
power spectra, as there is now a growing number of dedicated AGN
monitoring campaigns. One of the goals of these studies is to use the
characteristic timescales in the PDS to estimate the black hole mass in
AGN.  In particular we stress that the high-frequency break is {\em
not} constant, even {\em within} the low/hard state, so while this
break can be used as an {\em indicator} of black hole mass, it cannot
be used to make accurate predictions. We also stress that at high mass
accretion rates the GBH energy spectra are often dominated by the disc
emission This has very little variability compared to the Comptonized
tail (Churazov, Gilfanov \& Revnivtsev 2001), so it suppresses the rms
variability in the high/soft state GBH. The much lower disc temperature
in AGN means that their spectra are dominated by the Comptonized tail,
so this bandpass effect means that AGN can have much higher apparent
rms variability. This removes one of the problems in identifying the
often rapidly variable narrow line Seyfert 1 galaxies (Leighly 1999) as
the supermassive  counterparts of the high/soft state (Pounds et al.
1995), while other issues due to the shape of the PDS can be resolved
by recognizing the diversity of possible high mass accretion rate
states. We caution that Cyg X-1 does not sample a large range in mass
accretion rate, and shows a possibly unique PDS in its high/soft state,
so using this object as a template for AGN behaviour can severely
distort attempts to scale the variability between supermassive and
stellar black holes.

\begin{figure*}
\begin{center}
\leavevmode \epsfxsize=15.5cm \epsfbox{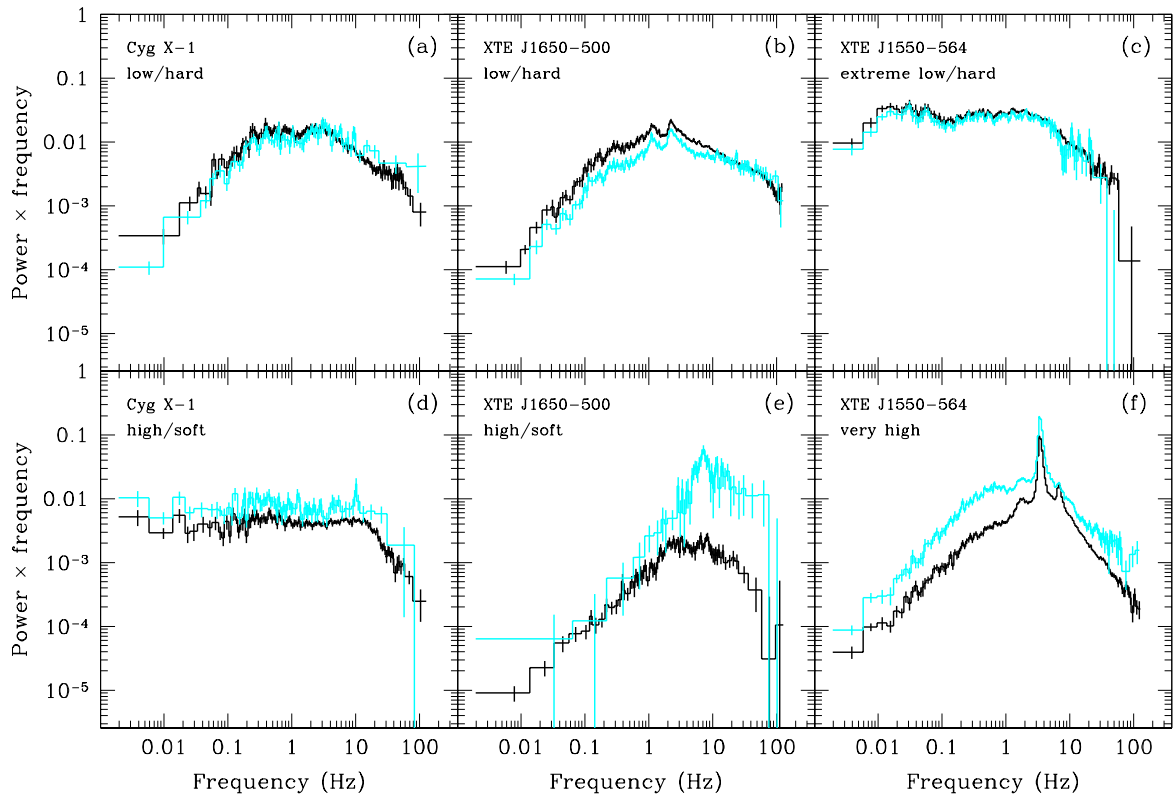}

\end{center}
\caption{Power density spectra from GBH over the full (2--60~keV:
black) and high-energy ($\sim$13--25~keV: grey) bandpass of the PCA.}
\label{fig:pds}
\end{figure*}

\begin{figure*}
\begin{center}
\leavevmode \epsfxsize=15.5cm \epsfbox{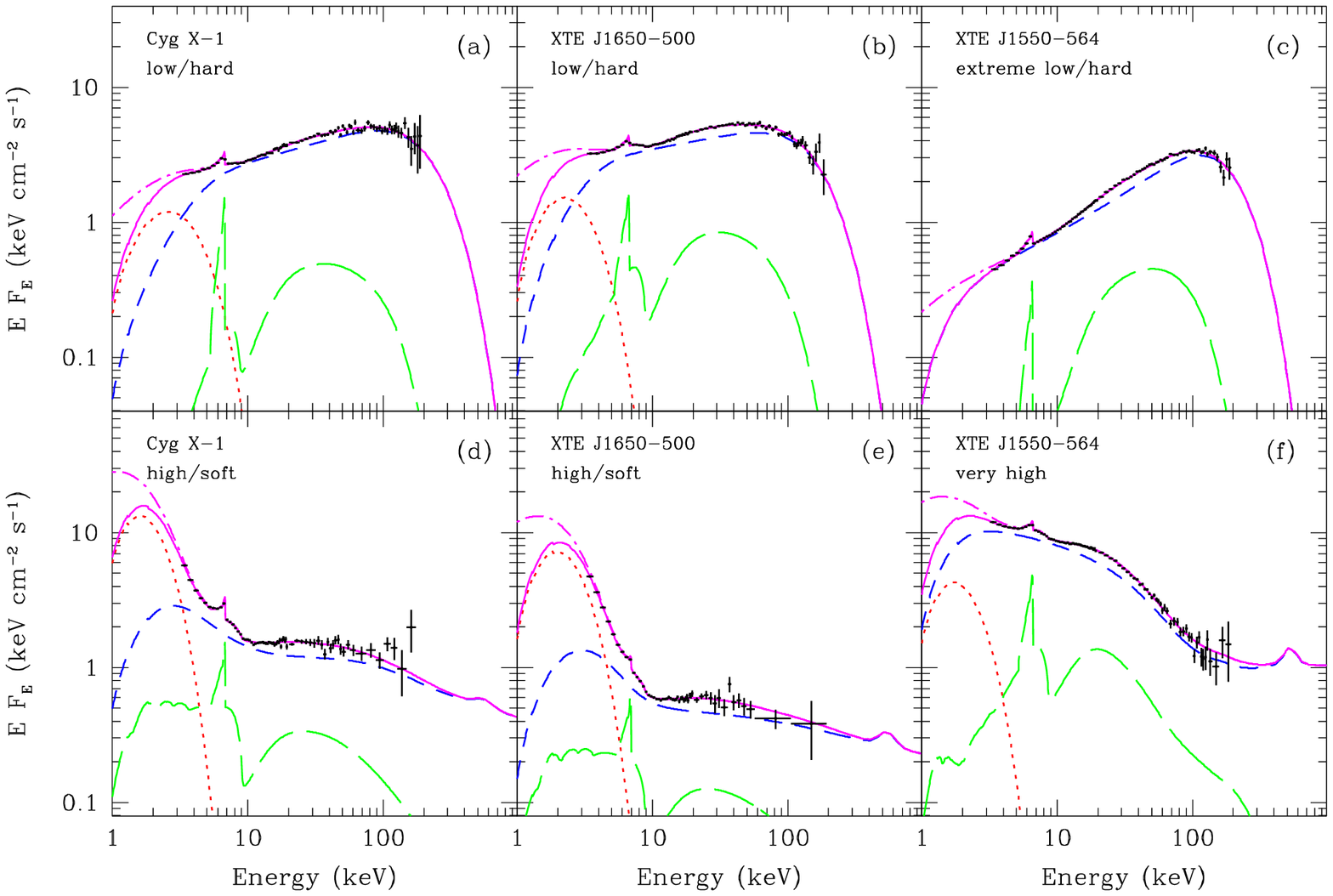}
\end{center}
\caption{Unfolded energy spectra from PCA and HEXTE and best-fitting
models corresponding to the each power spectrum shown in
Fig.~\ref{fig:pds}. The model components are plotted with different
curves: the multicolour disc (dotted), complex Comptonization (dashed)
and its reflection (long-dashed). The solid curve shows the sum. The
dash-dotted curve represents unabsorbed total model.} \label{fig:spec}
\end{figure*}

\section{Power spectra of GBH}

\subsection{The data}

We extract power density spectra of several GBH corresponding to
various spectral states in (1/256)--128~Hz frequency band from the
Proportional Counter Array (PCA) onboard \emph{Rossi X-ray Timing
Explorer} (\emph{RXTE}). These PDS, shown in Fig.~\ref{fig:pds} in
black, are calculated over the full energy band of the PCA (2--60~keV).
We also extract high-energy power spectra, from $\sim$ 13-25~keV
(starting at PCA channel 36 for Cyg X-1 and XTE J1550--564, and 26 for
XTE J1650--500, and ending at channel 71 to avoid the background
dominated energies), shown in Fig.~\ref{fig:pds} in light grey (cyan in
colour). This high-energy band excludes most of the disc emission, so
these PDS show the variability of the Comptonized tail alone.

We also extract both PCA and HEXTE energy spectra from the same
observations using standard extraction and background subtraction
criteria. These are shown in Fig.~\ref{fig:spec}, where they are
deconvolved from the instrument responses using the best-fitting model
consisting of a multicolour accretion disc ({\sc diskbb}: Mitsuda et
al.  1984), complex Comptonization ({\sc eqpair}: Coppi 1999) and its
reflection (Magdziarz \& Zdziarski 1995; plus {\sc diskline}: Fabian et
al. 1989). The total spectral model is also shown corrected for
interstellar absorption. The power and energy spectra in
Figs.~\ref{fig:pds} and \ref{fig:spec} were extracted from observations
with the following archival numbers a: 50110-01-38-01 ($L/L_{\rm Edd}$
= 0.011), b: 60113-01-04-00 (0.049), c: 80135-01-02-00 (0.033), d:
10512-01-09-00 (0.063), e: 60113-01-19-00 (0.060), and f:
30191-01-13-00 (0.25).  All $L/L_{\rm Edd}$ values are calculated using
bolometric luminosity estimated from the best-fitting model shown in
Fig.~\ref{fig:spec}. We used distances of 2 (Herrero et al. 1995), 5.3
(Orosz et al. 2002) and 4 kpc (Tomsick et al. 2003) and masses of 10
(Gierli{\'n}ski et al. 1999), 10 (Orosz et al. 2002) and 10 M$_{\odot}$
(Orosz et al. 2004) for Cyg X-1, XTE J1550--564 and XTE J1650--500,
respectively.

\subsection{Low/hard state}

Cyg X-1 is often used as the `canonical' GBH. Figs.~\ref{fig:pds}(a) and
\ref{fig:spec}(a) show a typical power and energy spectrum from its
low/hard state. The PCA energy spectrum is dominated by the hard
Comptonized component, with little emission from the soft component
visible in the data. The power spectra of the PCA data over the full-
and high-energy bandpass are remarkably similar, with integrated
fractional rms of $26.4\pm0.2$ and $24.8\pm0.5$ per cent, respectively,
showing that the major variability on these timescales is produced by
the normalization of the tail varying, rather than its spectral shape.
The typical low/hard state band-limited noise is evident in the PDS,
with most of the power contained between $\sim$ 0.2--3~Hz so that the
flat top spans an order of magnitude in frequency.

As opposed to the Cyg X-1, which is a persistent high-mass X-ray
binary, the low-mass X-ray binary XTE J1650--500 is a transient source.
Its low/hard state PDS is shown in Fig.~\ref{fig:pds}(b), with
corresponding energy spectrum in Fig.~\ref{fig:spec}(b). These data
were chosen so that the energy spectrum closely matched that of Cyg X-1
in Fig.~\ref{fig:spec}(a). The power spectra are similar, but differ in
detail. Firstly, the shape of the power spectrum is less smooth -- the
band-limited noise is more clearly associated with QPO components in
XTE J1650--500 than in Cyg X-1. Secondly the total rms is actually
slightly {\em decreased} at higher energies, with $19.8\pm 0.1$ per
cent compared to $24.5\pm 0.1$ in the full energy band, showing that
the Comptonizing tail is pivoting with changing soft photon input (see
Zdziarski et al. 2002 for similar behaviour in some `failed transition'
Cyg X-1 data).

Transient GBH allow us to probe the behaviour of the power spectra over
a much wider range of low/hard states than are sampled by Cyg X-1,
which actually varies rather little in total bolometric power (e.g.
Done \& Gierli{\'n}ski 2003). Fig.~\ref{fig:pds}(c) shows the power
spectra of an extreme low/hard state from another transient,
XTE~J1550--564, where the (approximately) flat part of the power
spectrum now extends over $\sim$~2.5 orders of magnitude in frequency.
The corresponding energy spectrum (Fig.~\ref{fig:spec}c) shows that the
Comptonized tail completely dominates the emission. The total rms
variability of the power spectra is unchanged between the full PCA and
high-energy PCA bandpasses, at $40.8\pm 0.4$ and $39.8\pm 0.7$ per
cent, respectively, showing that the variability is dominated by simply
the normalization of the tail, as in the Cyg X-1 low/hard data
(Fig.~\ref{fig:pds}a).

\begin{figure}
\begin{center}
\leavevmode \epsfxsize=8.5cm \epsfbox{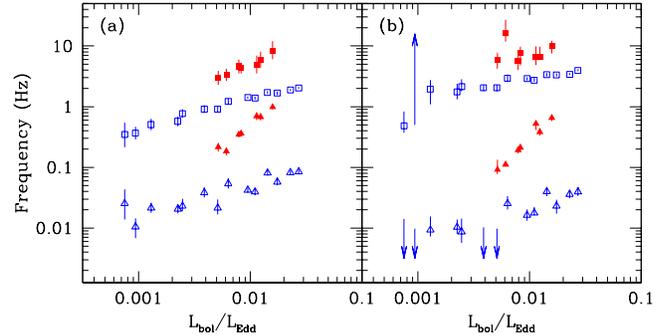}

\end{center}
\caption{Characteristic frequencies in the hard-state PDS of the 2000
(filled symbols) and 2002 (open symbols) outbursts of XTE J1550--564.
Triangles and squares correspond to lower and upper characteristic
frequency, respectively. Panel (a) shows the peak frequencies of two
Lorentzians, panel (b) break frequencies of the broken power law. In
the 2000 outburst the system reached the low/hard state from high/soft
or very high state, and the pattern of PDS changes reflects those
typically seen during the transition in that the low frequency break
moves by much more than the high-frequency break. The 2002 outburst
showed only low/hard state spectra, and here the low- and
high-frequency breaks are consistent with changing together as the
source dims.} \label{fig:freq}
\end{figure}

We can track the variation in these characteristic frequencies in the
power spectrum throughout the low/hard state in this transient system.
While multiple Lorentzian components are required to properly fit the
PDS shape (Nowak 2000; Belloni, Psaltis \& van der Klis 2003;
Pottschmidt et al. 2003), the broad shape can be fairly well
approximated by two Lorentzians (e.g. Axelsson, Borgonovo \& Larsson
2005). We fitted the XTE J1550--564 power spectra from its 2000 and
2002 outbursts where the source was in the low/hard state by two
Lorentzians, which gave a good description of the data. We also fitted
the PCA/HEXTE broad-band energy spectra with a simple model consisting
of the disc blackbody and thermal Comptonization (the model used by
Done \& Gierli{\'n}ski 2003) and used it to estimate the bolometric
luminosity as a fraction of Eddington luminosity. The resulting peak
frequencies of the two Lorentzians as a function of luminosity are
shown in Fig. \ref{fig:freq}(a).

However, doubly broken power laws are generally used to fit AGN data. We would
like to stress that generally these are not good description of the GBH power
spectra, where much better statistics reveals complex spectral shape, not
entirely consistent with a doubly broken power law. Nevertheless, we fitted the
XTE J1550--564 power spectra by this model for comparison. We fixed the low-
frequency PDS slope at zero, and the high-frequency slope at $-2$, and fitted
for the low- and high-frequency breaks and the `flat top' slope (which is
generally close to unity), together with the overall normalization. We show the
best-fitting break frequencies in Fig.~\ref{fig:freq}(b). Though the break
frequencies are different then the Lorentzian peak frequencies, the general
behaviour from both models is similar: {\em both} frequencies move. This is more
apparent for the low frequency break, which can change by at least a factor of
$\sim$~50 between the outbursts, while the high-frequency break can change by at
least a factor $\sim$~5 {\em from the same object}. This clearly shows that
characteristic frequencies in the low/hard state power spectra cannot be used as
an accurate tracer of black hole mass.

We stress that these data are all taken from {\em within} the low/hard
state, with energy spectral index $\alpha\sim$1.5--1.6 and are {\em
not} intermediate state or transition spectra.  However, the behaviour
seen in the 2000 outburst, where the low frequency break moves by much
more than the high-frequency break, continues into the hard-to-soft
transition. This has the effect of decreasing the extent of the flat
top and hence decreasing to the total rms (Gilfanov et al. 1999;
Churazov et al. 2001; Pottschmidt et al. 2003, Axelsson et al. 2005). A
very simple model for this correlated spectral and variability change
is one in which a hot inner flow filters a spectrum of fluctuations
from an outer cool disc. The inner and outer radius of the hot flow
determines the high- and low-frequency breaks, respectively. As the
disc moves inwards, the outer edge of the hot flow decreases, and
$\nu_{\rm low}$ increases (Churazov et al. 2001), while the increasing
disc flux for Compton scattering means that the spectrum gets softer.

\subsection{High/soft state}

Figs.~\ref{fig:pds}(d) and \ref{fig:spec}(d) show power and energy
spectra of a typical Cyg X-1 high/soft state. The energy spectrum is
now dominated by the disc component in the lowest PCA channels, but
with a tail to higher energies. The disc emission is rather constant,
while the tail is more variable (Churazov et al. 2001) so the
contribution of the disc to the PCA spectrum will clearly suppress the
rms variability observed. Fig.~\ref{fig:pds}(d) shows that the total
rms observed increases from $19.9\pm 0.5$ to $27.3\pm 1.2$ per cent
when the power spectrum is made from the high-energy data alone, where
there is negligible disc emission. The power spectral shape is also
clearly very different to that seen in the low/hard state. Its shape is
now similar to a low pass filter, with only a single characteristic
frequency which is the high-frequency rollover, $\nu_{\rm high} \sim
10$~Hz, while the flat top extends for over 4 decades in frequency.

Fig.~\ref{fig:pds}(e) shows the power spectrum from equivalent
high/soft state data from XTE J1650--500. These were chosen to closely
match the energy spectrum to that seen from the Cyg X-1 high/soft data
(Fig.~\ref{fig:spec}e). Despite the strong energy spectral
similarities, the power spectra are markedly {\em different} to that
from the high/soft state in Cyg X-1 (Fig.~\ref{fig:pds}d). The power
spectrum of the high/soft state of XTE J1650--500 is strongly peaked.
This matches smoothly onto the softest low/hard state power spectra
described above, where the flat part of the power spectrum shrinks to a
very small frequency range. Cyg X-1 {\em can} also show such peaked
power spectra (during its `failed' transitions to the high/soft state,
Pottschmidt et al. 2003), but its `canonical' high/soft power spectrum
seems dominated by a rather different power spectral component (see
Fig.~\ref{fig:pds}(d), also Fig 16 of Axelsson et al. 2005).

The spectrum of the high/soft state of XTE J1650--500 shows that the
disc spectrum dominates the low energy PCA bandpass, so can strongly
suppress the apparent variability. The Compton tail alone (high-energy
PCA bandpass) has rms variability of $26.2\pm 1.8$ per cent, whereas
including the lower energy, disc dominated channels gives total
variability of only $7.2\pm 0.2$ per cent.

It is rather difficult accumulate enough signal-to-noise in the power
spectra of the hard tail in the high/soft state of GBH, especially when
the disc has high temperature. Even if the tail is $\sim 10$ per cent
of the bolometric luminosity, its contribution to the total PCA count
rate is much smaller for a high-temperature disc than for a lower
temperature one due to the decreasing instrument response at higher
energies. However, we have examined all the available high-energy power
spectra from the outbursts of many GBH and have {\em never} seen a
high/soft state PDS which resembles that of the `canonical' high/soft
state derived from Cyg X-1. The high-energy power spectra are typically
complex or strongly peaked, and show a variable rms from $5-30$ per
cent.

Thus the Compton tail in the transient GBH shows a rather varied power
spectrum in the high/soft state.  It can be highly variable, with rms of
$\sim 30$ per cent, or it can be fairly constant (though still more
variable than the disc) with rms of $\sim 5$ per cent, but clearly there
is no well defined power spectral amplitude. However, where the tail has
high rms variability, as seen in Cyg X-1, its {\em shape} is very
different, showing strongly peaked noise which resembles QPO components
rather than the smooth low pass filter characteristic of Cyg X-1. The
only transient GBH convincingly show a flat PDS extending over more than
3 orders of magnitude in frequency are those showing extreme low/hard
energy spectra (Fig.~\ref{fig:pds}c).

\subsection{The very high state}

Figs.~\ref{fig:pds}(f) and \ref{fig:spec}(f) show the power and energy
spectra of the very high state of XTE J1550--564. The energy spectrum
clearly has a disc component, but the Comptonized emission is also
very strong, and dominates most of the PCA emission. The power
spectral shape, especially at high energies, is similar to the
low/hard state power spectra i.e. band-limited noise, except that the
QPO is {\em much} stronger and dominates the PDS (and makes the GBH
PDS impossible to fit with the broken power law model). The total rms
for the full- and high-energy bandpass are $23.3\pm 0.1$ and $32.3\pm
0.2$ per cent respectively, so clearly again the variability is more
marked by excluding the disc emission, but the effect is not so
drastic as in the high/soft state as the strong tail means the disc
dilution is not so important. {\em All} very high state power spectra
have high rms, and have this rather typical peaked noise shape, with
strong QPOs.

Very high luminosities, even exceeding $L_{\rm Edd}$, can be seen from
GRS~1915+105. While this source can show unique variability modes,
probably connected with its uniquely high accretion rate (Done,
Wardzi{\'n}ski \& Gierli{\'n}ski 2004), it spends about half of its
time in a rather stable very high state (class $\chi$: Belloni et al.
2000). PDS of these data show that they are very similar to those shown
here for XTE J1550--564, with variability increasing at higher energies
(Zdziarski et al. 2005).

\section{Application to AGN}

The GBH power and energy spectra shown in Figs~\ref{fig:pds} and
\ref{fig:spec} span the range of spectral states seen. However, they
also illustrate the well-- known lack of clean 1 to 1 correspondence
between spectral state and luminosity (e.g. van der Klis 2001;
McClintock \& Remillard 2005) as all except the very high state are
within a factor few from $L/L_{\rm Edd}\sim 0.03$. While in general the
low/hard state is seen at $L/L_{\rm Edd} < 0.05$, it can extend up to
$\sim 0.2$ on the rapid rise to outburst. As we have yet to clearly
understand the relation between GBH and AGN energy spectra, we use
$L/L_{\rm Edd}$ as a guide to AGN spectral state. Those with $L/L_{\rm
Edd}>0.2$ are likely to correspond to high mass accretion rate GBH i.e.
high/soft or very high state.

The much longer timescales involved mean that only a few AGN have been
monitored to the extent that power spectra can be derived. The power
spectra are also more complex to calculate than those from GBH as the
data are generally unevenly sampled. This gives a broad window function
which redistributes power over a wide range of frequencies, so
reconstructing the intrinsic power spectrum is analogous to deriving
the intrinsic energy spectrum from the broad spectral response of an
X-ray proportional counter. Techniques for this reconstruction
generally rely on Monte-Carlo simulations of the stochastic noise
properties of a given power spectral form through the specific uneven
sampling pattern corresponding to each AGN (Done et al. 1992; Uttley,
McHardy \& Papadakis 2002). These derived power spectra typically
become less well defined at the lowest frequencies, so the
uncertainties increase in the last decade in frequency of each
reconstruction.

\begin{figure*}
\begin{center}
\leavevmode \epsfxsize=12cm \epsfbox{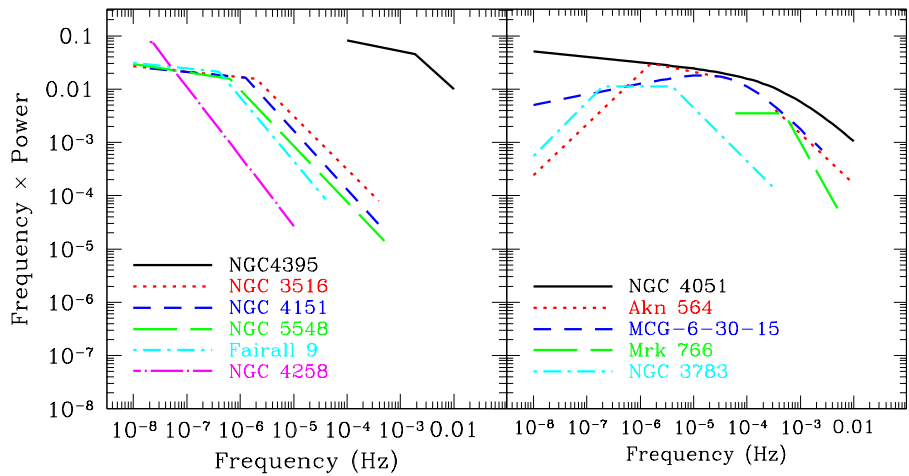}
\end{center}
\caption{A compilation of derived model power density spectra from AGN.
a) shows those for AGN with inferred $L/L_{\rm Edd}<0.2$, the
approximate maximum seen for the low/hard state in GBH. b) shows those
with $L/L_{\rm Edd} > 0.2$, so should correspond to high mass accretion
rate (high/soft and very high state) GBH. The PDS were taken from the
literature listed in the last column of Table \ref{tab:agn}.}
\label{fig:agn_pds}
\end{figure*}

Table ~\ref{tab:agn} gives basic data (black hole mass and Eddington
fraction, which are uncertain by factors of at least 2--3) of a
compilation of AGN from the literature which have published power
spectra, which we reproduce the best-fitting models in
Fig.~\ref{fig:agn_pds}a) and b) for low and high mass accretion rates,
respectively.

The different disc temperatures expected in AGN and GBH means that
there is a bandpass effect which masks the expected similarities in the
spectra and variability of the accretion flow. While the X-ray spectra
of both AGN and GBH should be dominated by the hard Comptonized tail at
low $L/L_{\rm Edd}$, the higher temperature disc expected in the GBH
means that this dominates their X-ray emission in the high/soft and
very high states, while the corresponding AGN spectra will be dominated
by the soft Comptonized tail seen in these states (see also McHardy et
al.  2004). Thus the AGN power spectra should be compared only to the
high-energy power spectra in GBH (cyan points in Fig~\ref{fig:pds})
{\em not} that derived from the total X-ray bandpass (black points).

However, there is also a secondary issue which is the {\em shape} of
the power spectrum.  Cyg X-1 is {\em not} representative of the
majority of the GBH in terms of its power spectra, despite being used
as the `canonical' object. Firstly, it is a persistent source,
spanning very little range in bolometric luminosity, so never shows a
very high state, nor an extreme low/hard state. Even over its limited
observed range in $L/L_{\rm Edd}$ it generally has a {\em different}
power spectra to those of other (transient) GBH with similar energy
spectra. In its high/soft state this is very marked, as the power
spectra of the Compton tail has approximately equal power per decade
between $10^{-3}$ and $10$~Hz (probably extending down below
$10^{-6}$~Hz: Reig, Papadakis \& Kylafis 2002), giving a high rms of
$\sim 30$ per cent.  By contrast the transient GBH with equivalent
energy spectra show Compton tail power spectra which are often sharply
peaked, or are complex, and have rms from $5-30$ per cent.  The
low/hard state power spectrum of Cyg X-1 is also subtly different from
those seen in the transient systems, in that the QPO features are
generally less marked.

\begin{table*}
\begin{tabular}{lcccccc}
\hline Source Name & $\log M_{\rm BH}/$M$_\odot$ & Method & $L/L_{\rm
Edd}$ & Type &
Ref.$^1$ & Ref.$^2$\\
\hline
NGC 4258 & 7.62 & megamaser & 0.005 & LLAGN & WU02 & MU05\\
Fairall 9 & 7.91 & reverberation & 0.16 & S1 & WU02 & M03\\
NGC 5548 & 8.03 & reverberation & 0.05 & S1 & WU02 & M03\\
NGC 4151 & 7.13 & reverberation & 0.03 & S1.5 & WU02 & M03\\
NGC 3516 & 7.36 & reverberation & 0.06 & S1 & WU02 & M03\\
NGC 4395 & 5.56($<$4.81) & reverberation(stellar velocity) &
0.0012($>$0.007) & S1 & P05, FH03 & VFI04\\

\hline
NGC 3783 & 6.94 & reverberation & 0.23 & S1 & WU02 & M03\\
Akn 564 & 6.9 & reverberation & 0.96 & NLS1 & R04 & V03, P02\\
MCG-6-30-15 & 6.71 & stellar velocity & 0.40 & NLS1 & M05 & M05\\
Mkn766 (aka NGC4253) & 6.54 & optical luminosity & 0.56 & NLS1 & WU02 & VF03\\
NGC 4051 & 6.13 & reverberation & 0.21 & NLS1 & WU02 & M04\\
\hline
\end{tabular}
\caption{A compilation showing estimated mass and $L/L_{\rm Edd}$ for
the AGN shown in Fig.~\ref{fig:agn_pds} (i.e. those with published
power spectra). The references are as follows: [FH03] Filippenko \& Ho
2003; [MU05] Markowitz \& Uttley 2005; [M03] Markowitz et al. 2003;
[M04] McHardy et al. 2004; [M05] McHardy et al. 2005; [P02] Papadakis
et al. 2002; [P05] Peterson et al. 2005; [R04] Romano et al. 2004;
[VF03] Vaughan \& Fabian 2003; [VFI04] Vaughan, Fabian \& Iwasawa 2004;
[V03] Vignali et al. 2003; [WU02] Woo \& Urry 2002. Ref.$^1$ refers to
the mass and luminosity estimate, Ref.$^2$ to the power spectrum shown
in Fig.~\ref{fig:agn_pds}.} \label{tab:agn}
\end{table*}

\subsection{Low $L/L_{\rm Edd}$ AGN}

A key goal of the AGN PDS studies is to use the high-frequency break as
a tracer of black hole mass. Fig.~\ref{fig:freq} shows that the
high-frequency break of a broken power law description of the low/hard
GBH PDS remains constant to within a factor $\sim$~5, so while this
cannot be used as an accurate tracer of black hole mass, it may give a
zeroth order indication.

The AGN shown in the left panel of  Fig.~\ref{fig:agn_pds} with
$L/L_{\rm Edd} < 0.20$ could correspond to a variety of spectral
states. Nonetheless, their PDS are consistent with being similar in
both normalization and shape, and the majority of AGN on this plot (all
with rather similar masses around $\sim 10^{7.5} M_\odot$ and accretion
rates of $L/L_{\rm Edd} \sim 0.1$: see Table~\ref{tab:agn}), are
compatible with scaling the high-frequency break from that seen in the
bright low/hard state of Cyg X-1.

However, there are two clear exceptions to this, both of which are
substantially lower $L/L_{\rm Edd}$ objects. NGC~4258 again has similar
mass of around $\sim 10^{7.5} M_\odot$, so its lower frequency break
could be due to it corresponding to a more extreme low/hard state (see
Fig~\ref{fig:freq}). This is a clear counterexample to the idea of a
universal PDS shape which depends only on mass (e.g Papadakis 2004;
O'Neill et al. 2005), as indeed is shown by Fig.~\ref{fig:freq}.

However, the lowest mass AGN known, NGC4395, is not so easy to fit into
this pattern. From the reverberation mass, it should have a
high-frequency break only a factor 100 higher than that from the other
objects. Yet its break is $\sim 4- 5$ orders of magnitude higher than
that for NGC~4258, the most obvious object to scale from in terms of
similarly low $L/L_{\rm Edd}$, and $\sim 3$ orders of magnitude from
the majority of AGN in the figure. Either there is truly a difference
in the variability properties of AGN and GBH, or the black hole mass in
NGC~4395 is overestimated. Some evidence for the latter possibility is
that this is the only object which has significantly different masses
inferred from reverberation and stellar velocity dispersion estimates
(refs). Assuming instead that the mass is $\sim 10^{4}M_\odot$,
consistent with the stellar velocity upper limit, this gives $L/L_{\rm
Edd}\sim 0.04$ and the PDS scales with mass in the same way as the
majority of AGN on the figure.

\subsection{High $L/L_{\rm Edd}$ AGN}

By contrast, Fig.~\ref{fig:agn_pds}b shows those AGN with $L/L_{\rm Edd} >
0.20$. Most of these have masses of $\sim 10^{6.5}M_\odot$ but their PDS are
very heterogeneous. This corresponds well with the multiple different types of PDS
shape which can be seen in the high/soft and very high states from GBH.

Two of these AGN (NGC 4051 and MCG-6-30-15, both NLS1's) clearly show flat noise
power extending over more than 4 decades in frequency. These look very like that
of the high/soft state in Cyg X-1, taking into account uncertainties in the low
frequency slope of the power spectra (McHardy et al. 2004; 2005).  The total rms
power in both is of order 25 per cent, again similar to that of the Comptonized
tail in the high/soft state of Cyg X-1 (McHardy et al. 2004). Thus these AGN can
be interpreted as the supermassive counterpart of the high/soft state Cyg X-1
data, which is {\em not} like the majority of transient GBH power spectra at
high accretion rates.

However, another NLS1, Akn564, convincingly shows {\em band-limited}
flat top noise, where the flat top extends over only $\sim$2 decades in
frequency (Fig.~\ref{fig:agn_pds} taken from a combination of low- and
high-frequency data in Markowitz et al. 2003 and Vignali et al. 2004).
Such band-limited, high rms ($\sim$ 30 per cent) noise is consistent
with the extreme low/hard state of GBH (Fig.~\ref{fig:pds}c). However,
Akn 564 has high $L/L_{\rm Edd}\sim 1$ (Table~\ref{tab:agn}), so should
instead be comparable to the high mass accretion rate GBH. Its PDS is
clearly {\em unlike} that of NGC~4051 (and the high/soft state of Cyg
X-1). Instead it could correspond to the PDS of the Compton tail seen
in the very high state (transient) GBH. These likewise show
band-limited noise similar to that of the low/hard state, with the flat
top extending over two decades in frequency, though this is usually
accompanied by a strong QPO (Fig.~\ref{fig:pds}f, Zdziarski et al.
2005). However, current AGN data probably cannot rule out the presence
of even a very strong QPO such as that seen in the very high state
since the uneven sampling strongly smears out the power (Vaughan \&
Uttley 2005).  If a QPO is not present with better sampling of the AGN
light curve then it may be that Akn 564 instead represents a scaled up
version of a very high state from a {\em persistent} GBH. There are as
yet no data on such GBH systems (Cyg X-1 never shows a very high
state), but we speculate by analogy with the Cyg X-1 low/hard state PDS
that these would have weaker QPO features than those seen in the
transient systems.

Band-limited noise is also probably detected in another high $L/L_{\rm
Edd}$ AGN, NGC~3783 (Markowitz et al. 2003).  Here the situation is
rather less clear, as $L/L_{\rm Edd}\sim 0.2$, i.e. close to the
maximum low/hard state luminosity. Uncertainties in mass and bolometric
luminosity mean this could easily fall or rise by a factor 2--3. Thus
this could either represent a bright low/hard state or a dim very high
state (sometimes termed an intermediate state: Belloni et al. 1996;
M{\'e}ndez \& van der Klis 1997). These different `states' merge
smoothly together in the GBH, which can be explained in the context of
a truncated disc extending closer to the black hole, increasing the
overlap between the disc and hot inner flow as the source moves from
low/hard to intermediate and very high states (Wilson \& Done 2001;
Kubota \& Done 2004).

The only other high $L/L_{\rm Edd}$ AGN with published power spectrum
is the NLS1, Mkn766. Here there is no data on long timescales, but the
(evenly sampled) XMM-Newton data strongly indicate a break at $\sim
5\times 10^{-4}$ Hz (Vaughan \& Fabian 2003; Vaughan, Fabian \& Iwasawa
2004). The normalization of the flat top noise is lower than that for
the other AGN (high or low $L/L_{\rm Edd}$), so this may correspond to
a high state counterpart of a transient GBH.

\section{The high/soft state of Cyg X-1 and its implications for AGN}

Despite its use a template PDS, Cyg X-1's timing properties are not
typical of GBH. The high/soft state PDS is markedly different, with
flat top (in $\nu P_\nu$) extending over at least four orders of
magnitude in frequency.  By contrast, power spectra of the
Compton tail in other GBH in the high/soft state are instead usually
strongly peaked at $\sim$10 Hz, or have complex shapes.

The difference in Cyg X-1 is perhaps connected to the persistent nature
of its emission due to the large accretion rate from its high-mass
companion OB star. The X-ray bolometric luminosity varies by very
little, $L/L_{\rm Edd}$ = 0.01--0.1, over timescales of weeks--years.
By contrast, most GBH have low-mass companions, with much lower mass
transfer rates through the Roche lobe so have transient outbursts where
the accretion flow luminosity changes from $L/L_{\rm Edd}\la 10^{-7}$
to 0.05--1 (e.g. McClintock, Narayan \& Rybicki 2004). We speculate
that the large dynamic range of variability seen in transients may
excite a rather different noise spectrum of turbulence in the disc. The
observed PDS is the convolution of this intrinsic power spectrum with
the response of the disc (e.g. Psaltis \& Norman 2000). For a
particular disc geometry, the response of the disc (its filter) is
fixed (Churazov et al. 2001; Psaltis \& Norman 2000), but if the
intrinsic noise spectrum changes then the observed PDS will also
change.

It is currently unclear whether the disc instability which causes the
dramatic outbursts in the transient GBH should also operate in AGN i.e.
whether the initial spectrum of fluctuations is similar to that in Cyg
X-1 or the transient GBH.  The instability is triggered by the dramatic
increase in opacity caused by the partial ionization of hydrogen at
temperatures of $\sim 10^4$~K.  Such temperatures are clearly expected
in AGN (Siemiginowska, Czerny \& Kostyunin 1996; Burderi, King \&
Szuszkiewicz 1998).  However, this needs to propagate globally through
the disc rather than being a purely local instability in order to
produce dramatic outbursts, which requires a {\em change} in the
effective $\alpha$ viscosity parameter. This is probably produced in GBH
discs by a change in viscosity mechanism, from the MRI where hydrogen is
ionized, to much weaker processes (eg spiral arms) where hydrogen is
neutral. Physically, this could be linked to the large change in density
of free electrons between the parts of the disc in which hydrogen is
ionized and neutral. If the neutral disc is {\em very} neutral, then the
lack of charge carriers suppresses the coupling of magnetic fields to
the disc, shutting off the MRI viscosity (Gammie \& Menou 1998).
However, AGN discs at the critical hydrogen ionization temperature are
much less dense than GBH discs at this temperature. Recombination
processes, especially three body recombination, will be much less
efficient at these lower densities, so while hydrogen is mainly neutral
there may be enough free electrons from potassium/iron etc to allow the
MRI to still operate as a viscosity mechanism (Menou \& Quataert 2001).
Thus the hydrogen ionization point would not trigger a global
instability in AGN discs (Janiuk et al. 2004). However, this is still to
some extent speculative, so it is not yet known whether we expect AGN
variability to be more like persistent GBH (i.e. Cyg X-1) or like the
transients.  The {\em observation} of a Cyg X-1 like high/soft power
spectrum in the nearby NLS1 NGC 4051 argues for a suppression of the
hydrogen ionization disc instability in this AGN, where the flat part of
the power spectrum extends over more than 4 decades in frequency
(McHardy et al. 2004). The lack of clear QPO signatures in AGN (while
mainly due to lack of statistics on the relevant timescales: e.g.
Vaughan 2005) may also be a feature of a different disc turbulence
spectrum characteristic of persistent sources. Better AGN power spectra,
with better constraints on potential QPO signatures could give
observational insights into the operation of the MRI and hydrogen
ionization instability in these massive discs.

\section{Conclusions}

Power spectra of AGN are used to estimate the black hole mass by
scaling characteristic timescales (breaks) to those seen in the
Galactic sources, most commonly using Cyg X-1 as the template PDS.
However, this assumes that the PDS shape is constant, while the GBH
show clearly that the characteristic frequencies change significantly
even within a single GBH state, and that the whole shape of the PDS can
change with state transitions. However, this does not mean that scaling
cannot yield useful constraints, though it certainly cannot give
accurate black hole masses. The high-frequency break changes by only a
factor $\sim$~5 within the low/hard state of GBH, and this is rather
less than the discrepancy inferred from relating break frequency to
mass for the lowest mass known AGN, NGC~4395. Either the mass of this
black hole is overestimated (for which there is some additional
evidence), or there is some other variability process occurring which
breaks the correspondence between this AGN and the GBH.

At high mass accretion rates, the X-ray spectra from GBH can be
dominated by the rather stable disc emission, while the much lower disc
temperature in AGN means that their X-ray spectra are dominated instead
by the Compton tail. This bandpass effect means that the low
variability often cited as being characteristic of the high/soft state
is characteristic of the {\em disc}. At higher energies, where
Comptonization dominates, GBH can exhibit large amplitudes of
variability, as required to match with AGN, especially the NLS1's. The
shape of the PDS in the high/soft state is generally strongly peaked,
not at all like the low pass filter PDS characteristic of the high/soft
state in Cyg X- 1. Since some NLS1s (NGC~4051 and MCG-6-30-15: McHardy
et al. 2004; 2005) show PDS which look convincingly like Cyg X-1, we
speculate that the Hydrogen ionization instability (which gives rise to
the transient behaviour in GBH) does not operate in AGN. However, there
are also some NLS1's which show band-limited PDS, similar to those seen
in the low/hard state. We show that such PDS can also be interpreted as
transient high/soft state PDS or very high state PDS, so that
band-limited noise is {\em not} a unique tracer of low mass accretion
rates.

AGN variability {\em can} be interpreted as the supermassive analogue of the GBH
variability, but only with careful matching of spectral states, and
careful consideration of the different spectral components.

%=================================================================

\section*{Acknowledgements}

We thank A. Markowitz, A. Siemiginowska and P. Uttley for enthusiastic
discussions and the anonymous referee for their useful comments. MG
acknowledges support through a PPARC PDRF.

\label{lastpage}


\begin{thebibliography}{}

\bibitem[\protect\citeauthoryear{Axelsson, Borgonovo, \&
Larsson}{2005}]{2005A&A...438..999A} Axelsson M., Borgonovo L., Larsson
S., 2005, A\&A, 438, 999

\bibitem[\protect\citeauthoryear{Belloni et
al.}{1996}]{1996ApJ...472L.107B} Belloni T., Mendez M., van der Klis
M., Hasinger G., Lewin W.~H.~G., van Paradijs J., 1996, ApJ, 472, L107


\bibitem[Belloni et al.(2000)]{2000A&A...355..271B} Belloni, T.,
Klein-Wolt, M., M{\' e}ndez, M., van der Klis, M., \& van Paradijs, J.\
2000, \aap, 355, 271

\bibitem[Belloni et al.(2002)]{2002ApJ...572..392B} Belloni, T., Psaltis,
D., \& van der Klis, M.\ 2002, \apj, 572, 392

\bibitem[Boroson(2002)]{2002ApJ...565...78B} Boroson, T.~A.\ 2002, \apj,
565, 78

\bibitem[Burderi et al.(1998)]{1998ApJ...509...85B} Burderi, L., King,
A.~R., \& Szuszkiewicz, E.\ 1998, \apj, 509, 85

\bibitem[Churazov et al.(2001)]{2001MNRAS.321..759C} Churazov, E.,
Gilfanov, M., \& Revnivtsev, M.\ 2001, \mnras, 321, 759

\bibitem[Collin \& Kawaguchi(2004)]{2004A&A...426..797C} Collin, S., \&
Kawaguchi, T.\ 2004, \aap, 426, 797

\bibitem[Coppi(1999)]{1999ASPC..161..375C} Coppi, P.~S.\ 1999, ASP
Conf.~Ser.~161: High Energy Processes in Accreting Black Holes, 161, 375

\bibitem[Cui et al.(1997)]{1997ApJ...484..383C} Cui, W., Zhang, S.~N.,
Focke, W., \& Swank, J.~H.\ 1997, \apj, 484, 383

\bibitem[Done \& Gierli{\' n}ski(2003)]{2003MNRAS.342.1041D} Done, C., \&
Gierli{\' n}ski, M.\ 2003, \mnras, 342, 1041

\bibitem[Done et al.(2004)]{2004MNRAS.349..393D} Done, C., Wardzi{\' n}ski,
G., \& Gierli{\' n}ski, M.\ 2004, \mnras, 349, 393

\bibitem[\protect\citeauthoryear{Done et al.}{1992}]{1992ApJ...400..138D}
Done C., Madejski G.~M., Mushotzky R.~F., Turner T.~J., Koyama K.,
Kunieda H., 1992, ApJ, 400, 138

\bibitem[Fabian et al.(1989)]{1989MNRAS.238..729F} Fabian, A.~C., Rees,
M.~J., Stella, L., \& White, N.~E.\ 1989, \mnras, 238, 729

\bibitem[Falcke et al.(2004)]{2004A&A...414..895F} Falcke, H., K{\"
o}rding, E., \& Markoff, S.\ 2004, \aap, 414, 895

\bibitem[\protect\citeauthoryear{Filippenko \&
Ho}{2003}]{2003ApJ...588L..13F} Filippenko A.~V., Ho L.~C., 2003, ApJ,
588, L13

\bibitem[Gammie \& Menou(1998)]{1998ApJ...492L..75G} Gammie, C.~F., \&
Menou, K.\ 1998, \apjl, 492, L75

\bibitem[Gierli{\' n}ski et al.(1999)]{1999MNRAS.309..496G} Gierli{\'
n}ski, M., Zdziarski, A.~A., Poutanen, J., Coppi, P.~S., Ebisawa, K., \&
Johnson, W.~N.\ 1999, \mnras, 309, 496

\bibitem[Gilfanov et al.(1999)]{1999A&A...352..182G} Gilfanov, M.,
Churazov, E., \& Revnivtsev, M.\ 1999, \aap, 352, 182

\bibitem[\protect\citeauthoryear{Herrero et
al.}{1995}]{1995A&A...297..556H} Herrero A., Kudritzki R.~P., Gabler
R., Vilchez J.~M., Gabler A., 1995, A\&A, 297, 556

\bibitem[\protect\citeauthoryear{Janiuk et al.}{2004}]{2004ApJ...602..595J}
Janiuk A., Czerny B., Siemiginowska A., Szczerba R., 2004, ApJ, 602, 595

\bibitem[]{k04} Kalemci, E.; Tomsick, J. A.; Rothschild, R. E.; Pottschmidt, K.;
Kaaret, P. 2004, ApJ, 603, 231

\bibitem[\protect\citeauthoryear{Kubota \&
Done}{2004}]{2004MNRAS.353..980K} Kubota A., Done C., 2004, MNRAS, 353,
980

\bibitem[Leighly(1999)]{1999ApJS..125..297L} Leighly, K.~M.\ 1999, ApJS,
125, 297

\bibitem[]{mz95}Magdziarz P., Zdziarski A., 1995, MNRAS, 273, 837

\bibitem[\protect\citeauthoryear{Markowitz \&
Uttley}{2005}]{2005ApJ...625L..39M} Markowitz A., Uttley P., 2005, ApJ,
625, L39

\bibitem[Markowitz et al.(2003)]{2003ApJ...593...96M} Markowitz, A., et
al.\ 2003, \apj, 593, 96

\bibitem[McHardy et al.(2004)]{2004MNRAS.348..783M} McHardy, I.~M.,
Papadakis, I.~E., Uttley, P., Page, M.~J., \& Mason, K.~O.\ 2004, \mnras,
348, 783

\bibitem[\protect\citeauthoryear{McHardy et
al.}{2005}]{2005MNRAS.359.1469M} McHardy I.~M., Gunn K.~F., Uttley P.,
Goad M.~R., 2005, MNRAS, 359, 1469

\bibitem[]{mr05} McClintock J.E., Remillard R., 2005, in "Compact Stellar X-ray
Sources," eds. W.H.G. Lewin and M. van der Klis, Cambridge University
Press, in press

\bibitem[McClintock et al.(2004)]{2004ApJ...615..402M} McClintock, J.~E.,
Narayan, R., \& Rybicki, G.~B.\ 2004, \apj, 615, 402

\bibitem[\protect\citeauthoryear{Mendez \& van der
Klis}{1997}]{1997ApJ...479..926M} M{\'e}ndez M., van der Klis M., 1997,
ApJ, 479, 926


\bibitem[Menou \& Quataert(2001)]{2001ApJ...552..204M} Menou, K., \&
Quataert, E.\ 2001, \apj, 552, 204

\bibitem[Mitsuda et al.(1984)]{1984PASJ...36..741M} Mitsuda, K., et al.\
1984, PASJ, 36, 741

\bibitem[Merloni et al.(2003)]{2003MNRAS.345.1057M} Merloni, A., Heinz, S.,
\& di Matteo, T.\ 2003, \mnras, 345, 1057

\bibitem[Nandra et al.(1997)]{1997ApJ...476...70N} Nandra, K., George,
I.~M., Mushotzky, R.~F., Turner, T.~J., \& Yaqoob, T.\ 1997, \apj, 476, 70

\bibitem[\protect\citeauthoryear{Nowak}{2000}]{2000MNRAS.318..361N} Nowak
M.~A., 2000, MNRAS, 318, 361

\bibitem[O'Neill et al.(2005)]{2005MNRAS.tmp..251O} O'Neill, P.~M., Nandra,
K., Papadakis, I.~E., \& Turner, T.~J.\ 2005, \mnras, 251

\bibitem[\protect\citeauthoryear{Orosz et al.}{2002}]{2002ApJ...568..845O}
Orosz J.~A., et al., 2002, ApJ, 568, 845

\bibitem[\protect\citeauthoryear{Orosz et al.}{2004}]{2004ApJ...616..376O}
Orosz J.~A., McClintock J.~E., Remillard R.~A., Corbel S., 2004, ApJ,
616, 376

\bibitem[\protect\citeauthoryear{Papadakis}{2004}]{2004MNRAS.348..207P}
Papadakis I.~E., 2004, MNRAS, 348, 207

\bibitem[Papadakis et al.(2002)]{2002A&A...382L...1P} Papadakis, I.~E.,
Brinkmann, W., Negoro, H., \& Gliozzi, M.\ 2002, \aap, 382, L1

\bibitem[Pottschmidt et al.(2003)]{2003A&A...407.1039P} Pottschmidt, K., et
al.\ 2003, \aap, 407, 1039

\bibitem[Pounds et al.(1995)]{1995MNRAS.277L...5P} Pounds, K.~A., Done, C.,
\& Osborne, J.~P.\ 1995, \mnras, 277, L5

\bibitem[]{pn00} Psaltis D., Norman C., 2000, astro-ph/0001391.

\bibitem[Reig et al.(2002)]{2002A&A...383..202R} Reig, P., Papadakis, I.,
\& Kylafis, N.~D.\ 2002, \aap, 383, 202

\bibitem[\protect\citename{Shakura} 1973]{1973A&A....24..337S} Shakura
N.~I., Sunyaev R.~A., 1973, A\&A,  24, 337

\bibitem[Siemiginowska et al.(1996)]{1996ApJ...458..491S} Siemiginowska,
A., Czerny, B., \& Kostyunin, V.\ 1996, \apj, 458, 491

\bibitem[]{tl95} Tanaka Y., Lewin W. H. G.  1995, in X--Ray Binaries,
ed. W. H. G. Lewin, J. van Paradijs \& E. van den Heuvel (Cambridge:
Cambridge Univ. Press), 126

\bibitem[\protect\citeauthoryear{Tomsick et
al.}{2003}]{2003ApJ...592.1100T} Tomsick J.~A., Kalemci E., Corbel S.,
Kaaret P., 2003, ApJ, 592, 1100

\bibitem[Uttley et al.(2002)]{2002MNRAS.332..231U} Uttley, P., McHardy,
I.~M., \& Papadakis, I.~E.\ 2002, \mnras, 332, 231

\bibitem[]{vk00} van der Klis M., 2000, ARA\&A, 38, 717

\bibitem[\protect\citeauthoryear{van der Klis}{2001}]{2001ApJ...561..943V}
van der Klis M., 2001, ApJ, 561, 943

\bibitem[\protect\citeauthoryear{Vaughan}{2005}]{2005A&A...431..391V}
Vaughan S., 2005, A\&A, 431, 391

\bibitem[\protect\citeauthoryear{Vaughan \&
Fabian}{2003}]{2003MNRAS.341..496V} Vaughan S., Fabian A.~C., 2003,
MNRAS, 341, 496

\bibitem[]{vu05}Vaughan S., Uttley P., 2005, preprint, astro-ph/0506456

\bibitem[\protect\citeauthoryear{Vaughan, Fabian, \&
Iwasawa}{2004}]{2004astro.ph.12695V} Vaughan S., Fabian A.~C., Iwasawa
K., 2004, preprint, astro-ph/0412695

\bibitem[\protect\citeauthoryear{Vignali et
al.}{2004}]{2004MNRAS.347..854V} Vignali C., Brandt W.~N., Boller T.,
Fabian A.~C., Vaughan S., 2004, MNRAS, 347, 854

\bibitem[Wilson \& Done (2001)]{wd01} Wilson C. D., Done C. 2001, MNRAS, 325, 167

\bibitem[White et al.(1984)]{1984A&A...133L...9W} White, N.~E., Fabian,
A.~C., \& Mushotzky, R.~F.\ 1984, \aap, 133, L9

\bibitem[\protect\citeauthoryear{Woo \& Urry}{2002}]{2002ApJ...579..530W}
Woo J.-H., Urry C.~M., 2002, ApJ, 579, 530

\bibitem[Zdziarski \& Gierli{\' n}ski(2004)]{2004PThPS.155...99Z}
Zdziarski, A.~A., \& Gierli{\' n}ski, M.\ 2004, Progress of Theoretical
Physics Supplement, 155, 99

\bibitem[Zdziarski et al.(2002)]{2002ApJ...578..357Z} Zdziarski, A.~A.,
Poutanen, J., Paciesas, W.~S., \& Wen, L.\ 2002, \apj, 578, 357

\bibitem[\protect\citeauthoryear{Zdziarski et
al.}{2005}]{2005MNRAS.360..825Z} Zdziarski A.~A., Gierli{\' n}ski M.,
Rao A.~R., Vadawale S.~V., Miko{\l}ajewska J., 2005, MNRAS, 360, 825

\end{thebibliography}
\end{document}